\documentclass{aa}

\newcommand{\msun}{$M_{\odot}\ $}

\begin{document}

\title{A study of accretion discs around rapidly rotating neutron 
stars in general relativity and its applications to four 
Low Mass X--ray Binaries}

\titlerunning{Accretion discs around rapidly rotating neutron stars}

\author{Sudip Bhattacharyya\inst{1,2}}

\authorrunning{S. Bhattacharyya}

\offprints{Sudip Bhattacharyya}

\institute{Joint Astronomy Program, Indian Institute of Science,
Bangalore 560012, INDIA \\
\and
Indian Institute of Astrophysics,
Bangalore 560 034, INDIA\\
(sudip@physics.iisc.ernet.in; sudip@iiap.ernet.in)}

\date{}

\abstract{We calculate the accretion disc 
temperature profiles, disc luminosities and 
boundary layer luminosities for rapidly rotating neutron stars considering the 
full effect of general relativity. We compare the theoretical values of 
these quantities with their values inferred from {\it EXOSAT} data for 
four low mass X--ray binary sources: XB 1820-30, GX 17+2, GX 9+1 and GX 349+2 
and constrain the values of several properties of these sources. According 
to our calculations, the neutron stars in GX 9+1 and GX 349+2 are rapidly 
rotating and stiffer equations of state are unfavoured.
\keywords{accretion, accretion discs --- relativity ---
stars: neutron --- stars: rotation --- X--rays: binaries}}

\maketitle

\section{Introduction}

A low mass X--ray binary (LMXB) is believed to contain either a weakly magnetised
neutron star or a black hole as the central accretor. The X--ray emission
arises from the innermost region of the accretion disc around the compact
star. In the case of a neutron star, there is an additional X--ray component
coming from the boundary layer of the star. Mitsuda et al. (1984) showed 
that the spectrum of a luminous LMXB can be fitted by the sum of a single
temperature blackbody spectrum (believed to come from the boundary layer)  
and a multicolour blackbody spectrum (may be originated from the accretion 
disc). However these authors used Newtonian models to fit the observed 
spectra. But near the surface of a neutron star, 
the accretion flow is expected to be governed by the laws of 
general relativity due to the presence of strong gravity. Therefore
general relativistic models should be used for the 
purpose of fitting to get the correct best-fit values of the parameters.
Besides, the principal motivation behind the study of the spectral and 
temporal behaviours of neutron star LMXBs is to understand the properties 
of very high $(\sim 10^{15}$ g cm$^{-3})$ density matter at the neutron star
core (van der Klis 2000). 
This is a fundamental problem of physics, which can not be addressed
by any kind of laboratory experiment. The only way to answer this question
is to assume an equation of state (EOS) model for the neutron star core, to 
calculate the structure parameters of the neutron star and hence to 
calculate an appropriate spectral model. By fitting such models (for different
chosen EOSs) to the observed data, one can hope to constrain the existing 
EOS models and hence to understand the properties of high density matter.
However, general relativistic calculation is essential to calculate the 
structure parameters of a neutron star and therefore to constrain the EOS 
models.

It is expected that the neutron stars in LMXBs are rapidly rotating due to 
accretion-induced angular momentum transfer. 
LMXBs are thought to be the progenitors of milli-second (ms) radio pulsars
(Bhattacharya \& van den Heuvel 1991) like PSR 1937+21 with $P \sim 
1.56$~ms (Backer et al. 1982). The recent discovery of ms $(P \sim 2.49$~ms) 
X--ray pulsations in XTE J1808-369 (Wijnands \& van der Klis 1998) has 
strengthened this hypothesis. Therefore it is necessary to
calculate the structure of a rotating neutron star considering the full effect 
of general relativity. This was done by Cook, Shapiro \& Teukolsky 
(1994) and the same 
procedure was used by Thampan \& Datta (1998), 
to calculate the luminosities of the disc and the boundary layer.

The disc temperature profile for a rapidly rotating neutron star was first
calculated by Bhattacharyya et al. (2000). These authors also compared their 
theoretical results with the {\it EXOSAT} data (analysed by White, Stella \&
Parmar 1988) to constrain different properties of the LMXB source Cygnus X-2.
The present work is a continuation of theirs, in which we constrain several 
properties of four LMXB sources: XB 1820-30, GX 17+2, GX 9+1 and GX 349+2, 
using the same procedure.
These sources were also observed by {\it EXOSAT} and the data were analysed 
by White et al. (1988).

XB 1820-30 is an atoll source which shows type I X--ray bursts. GX 17+2 and 
GX 349+2 are Z sources, of which the former shows X--ray bursts. GX 9+1 is an
atoll source. As all of them are LMXBs (van Paradijs 1995), the magnetic 
field of the neutron stars are believed to be decayed to lower values 
$(\sim 10^8$ G; see Bhattacharya \& Datta 1996 and Bhattacharya \& 
van den Heuvel 1991). Therefore, we ignore the effect of the 
magnetic field on the accretion disc structure in our calculations.

In section 2, we give the formalism of the work. We present the results 
and discussion in section 3 and give a summary in section 4.

\section{Formalism}

\subsection{Theoretical formulae}

In order to calculate the disc temperature profile, the disc luminosity 
and the boundary layer luminosity for a rapidly rotating neutron star 
considering the full effect of general relativity, we need to compute the 
structure of the rotating star. To do this, following Cook et al. (1994),
we choose a stationary, axisymmetric, asymptotically flat and 
reflection-symmetric (about the equatorial plane) metric,
given by
\begin{eqnarray}
dS^2 & = & g_{\rm \mu\nu} dx^{\rm \mu} dx^{\rm \nu}
(\mu, \nu = 0, 1, 2, 3)
\nonumber\\
 & = & -e^{\rm {\gamma + \rho}} dt^2 + e^{\rm {2\alpha}}
(d{\bar r}^2 + {\bar r}^2 d {\theta}^2) \nonumber \\
 &  &
 + e^{\rm {\gamma - \rho}} {\bar r}^2 \sin^2\theta
 {(d\phi - \omega dt)}^2 ~,
\label{eq: metric}
\end{eqnarray}
where the metric coefficients $\gamma, \rho, \alpha$ and the
angular speed ($\omega$) of zero-angular-momentum-observer
(ZAMO) with respect to infinity, are all functions of the
quasi-isotropic radial coordinate ($\bar r$) and polar angle
($\theta$). The quantity $\bar r$ is related to the
Schwarzschild-like radial coordinate ($r$) by the equation
$r = \bar r e^{\rm {(\gamma - \rho)/2}}$.
We use the geometric units $c = G = 1$ in this paper.

With the
assumption that the star is rigidly rotating and a perfect
fluid, we solve Einstein's field equations and the equation of
hydrostatic equilibrium (the equations are given in the 
appendix) self-consistently and numerically
from the centre of the star upto infinity
to obtain the metric coefficients and $\Omega_{\rm *}$
(angular speed of neutron star with respect to infinity)
as functions of $\bar r$ and $\theta$. The inputs of this
calculation are a chosen EOS and assumed values of the central density and
the ratio of polar to equatorial radii. The outputs are bulk structure 
parameters, such as gravitational mass ($M$),
equatorial radius ($R$), angular momentum ($J$), moment of inertia 
($I$) etc. of the neutron star. We can also calculate
the specific disc luminosity ($E_{\rm D}$),
the specific boundary layer luminosity ($E_{\rm BL}$),
the radius ($r_{\rm orb}$)
of the innermost stable circular orbit (ISCO) and specific
energy ($\tilde E$), specific angular momentum ($\tilde l$) \& angular
 speed ($\Omega_{\rm K}$) of a test particle in a Keplerian orbit (see
Thampan \& Datta 1998
for a description of the method of calculation).

Then we calculate the effective temperature of a geometrically 
thin blackbody disc, which is given by
\begin{eqnarray}
T_{\rm eff}(r) & = & (F(r)/\sigma)^{1/4} \label{eq: teff}
\end{eqnarray}
where $\sigma$ is the Stefan-Boltzmann constant and $F$ is
the X--ray energy flux (due to viscous dissipation) per unit surface 
area. We calculate
$F$ using the expression of  Page \& Thorne (1974):
\begin{eqnarray}
F(r) & = & \frac{\dot{M}}{4 \pi r} q(r)
\end{eqnarray}
\noindent where
\begin{eqnarray}
q(r) & = & -\Omega_{{\rm K},r} (\tilde{E} - \Omega_{\rm K}
\tilde{l})^{-2} \int_{r_{\rm in}}^{r} (\tilde{E} - \Omega_{\rm K}
\tilde{l}) \tilde{l}_{,r} dr
\label{eq: fr}
\end{eqnarray}
Here $r_{\rm in}$ is the disc inner edge radius and a
comma followed by a variable as subscript to a quantity,
represents a derivative of the quantity with respect to
the variable. 
The values for $\tilde{E}$ and $\tilde{l}$ are given by the
two conditions (circularity and extremum) for orbits 
(see Thampan \& Datta 1998; Bhattacharyya et al. 2000):

\begin{eqnarray}
\tilde{E} - \omega \tilde{l} & = & \frac{e^{(\gamma+\rho)/2}}{\sqrt{1-v^2}} \\
\tilde{l} & = & \frac{v \bar{r} e^{(\gamma-\rho)/2}}{\sqrt{1-v^2}}
\end{eqnarray}

\noindent where $v=(\Omega-\omega) \bar{r} e^{-\rho} sin \theta$ is the
physical
velocity of the matter.  The equation of motion in the azimuthal direction
and that in the time direction yield the Keplerian angular speed as
\begin{eqnarray}
\Omega_{\rm K} & = & e^{2\rho(\bar{r})}
\frac{\tilde{l}/ \bar{r}^2}{(\tilde{E}-\omega \tilde{l})} + \omega(\bar{r})
\end{eqnarray}

Eq. (4), that is exactly valid for a black hole accretor, is also valid
for a neutron star accretor, if the disc does not touch the star.
But, for a disc that extends upto the surface of the neutron star,
the torque at the disc inner edge will not vanish, and eq. (4) will
not be strictly valid for such a case.
However, for a very rapidly rotating neutron star, the angular speed 
of a particle at the disc inner edge will be close to the spin rate
of the star, and hence the torque is expected to be negligible.
According to Bhattacharyya et al. (2000), 
even when the spin rate is not large, eq. (4) should give approximately
correct results. This is because, Page \& Thorne (1974) argued
that the error in the calculation of $T_{\rm eff}$ will not be
substantial outside a radial distance $1.1 r_{\rm in}$. In our calculation, 
we find that almost whole of the disc X--ray flux comes from well-outside
this radial distance.

Eq. (\ref{eq: teff})  gives the effective disc temperature $T_{\rm eff} (r)$
with respect to an observer comoving with the disc. For our purpose, 
this expression of the temperature must be changed to that in the observer's 
frame, taking into account the
gravitational redshift and the rotational Doppler effect.
In order to keep our analysis tractable, we use
the expression given in Hanawa (1989) for this modification :
\begin{eqnarray}
1+z = (1-\frac{3M}{r})^{-1/2}
\end{eqnarray}
With this correction for $(1+z)$, we define a temperature relevant for
observations ($T_{\rm obs}^{\rm max}$) as:
\begin{eqnarray}
T_{\rm obs}^{\rm max} = \frac{1}{1+z} T_{\rm eff}^{\rm max} \label{eq: tobs}
\end{eqnarray}
\noindent where the superscript `max' denotes the maximum temperature value in 
the profile. $T_{\rm obs}^{\rm max}$ is related to the colour temperature 
$(T_{\rm col}^{\rm max})$ by $T_{\rm obs}^{\rm max} = 
T_{\rm col}^{\rm max}/f$, where $f$ is the colour factor (Shimura 
\& Takahara 1995; see also Bhattacharyya et al. 2000 for details). We 
compare the observationally inferred value of the maximum temperature with 
$T_{\rm col}^{\rm max}$.
It is to be remembered that eq. (8) is valid for a non-rotating neutron
star and a `face-on' (i.e., inclination angle $i = 0)$ observation. However, the
error involved by these assumptions is expected to be well-within the error
bars considered in subsection 2.3.

The structure of a neutron star for a given EOS is described
uniquely by two parameters : the  gravitational mass ($M$) and
the angular speed ($\Omega_{\rm *}$).  For each adopted EOS, we construct
constant $M$ equilibrium sequences with $\Omega_{\rm *}$ varying from the
non-rotating case (static limit) upto the centrifugal mass shed limit (rotation
rate at which inwardly directed gravitational forces are balanced
by outwardly directed centrifugal forces). So we are able to calculate 
$T_{\rm eff}^{\rm max}$, $E_{\rm D}$ and $E_{\rm BL}$ as functions of 
$M$ and $\Omega_{\rm *}$ for a chosen EOS model.

\subsection{Equations of State}

The neutron star structure parameters are quite sensitive to
the chosen EOS models. For the purpose of a general study, 
we have used the same four EOSs as considered in Bhattacharyya et al. (2000), 
namely, (A) Pandharipande (1971), 
(B) Baldo, Bombaci \& Burgio (1997), 
(C) Walecka (1974) and (D) Sahu, Basu \& Datta (1993).
EOS model (A) is for hyperonic matter. It is assumed that hyperonic 
potentials are similar to the nucleon--nucleon potentials, but altered 
suitably to represent the different isospin states. Model (B) is a microscopic
EOS for asymmetric nuclear matter, derived from the Brueckner--Bethe--Goldstone
many--body theory with explicit
three--body terms. The three--body force parameters are adjusted to give a
reasonable saturation point for nuclear matter. EOS model (C) corresponds to
pure neutron matter and is based on a mean--field theory with exchange of
scaler and (isoscalar) vector mesons representing the nuclear interaction.
Model (D) is a field
theoretical EOS for neutron--rich matter in beta equilibrium based on the
chiral sigma model. The model includes an isoscalar vector field generated
dynamically and reproduces the empirical values of the nuclear matter
saturation density and binding energy and also the isospin symmetry
coefficient for asymmetric nuclear matter.
Of these, model (A) is soft, (B) is intermediate in stiffness and
(C) \& (D) are stiff EOSs, with (D) as the stiffest. 

\subsection{Constraining procedure}

We choose four LMXB sources observed by {\it EXOSAT} 
(data analysed by White et al. 1988).
For each source, we take the best-fit values of the parameters  
$T_{\rm col}^{\rm max}$, $L_{\rm D}$ and $L_{\rm BL}$, where $L$ denotes 
the luminosity. On the other hand, for the chosen values of $f$, 
the accretion rate ($\dot M$) and $M$, we theoretically 
calculate the values of $T_{\rm col}^{\rm max}$, 
$L_{\rm D}$ and $L_{\rm BL}$ as functions of $\Omega_{\rm *}$, for an 
assumed EOS model. 
Then comparing these theoretical values with the observed ones, 
we constrain different parameters of the chosen source
(see Bhattacharyya et al. 2000 for detailed description).
However, to take into account 
the uncertainties in the fitting procedure and in the value of $z$, and
also those arising due to the simplicity of the model, we consider a range of
acceptable
values for $T_{\rm col}^{\rm max}$, $L_{\rm D}$ and $L_{\rm BL}$.
We take two combinations of deviations around the best-fit values, namely,
($10\%$, $25\%$) and ($20\%$, $50\%$), where the first number in parentheses
corresponds to the error in $T_{\rm col}^{\rm max}$ and the second to
the error in the best-fit luminosities.

\section{Results and Discussion}

In this paper, we calculate gravitational mass sequences for different
EOS models and constrain several properties of four LMXB sources. 
For the neutron star in each of the sources, we assume 
$M = 1.4$~\msun (i.e., the canonical mass value for neutron stars).
We take two values for $\cos i$ $(i$ is the inclination angle of the source)
for each source, namely, $0.2$ and $0.8$. These two widely 
different values ensure the sufficient generality of our results.
For the four sources, the best-fit values (White et al. 1988) of the parameters 
$T_{\rm col}^{\rm max}$, $L_{\rm D}$ and $L_{\rm BL}$ are 
given in Table 1.

We take the distance $(D)$ of the source XB 1820-30 as 6.4 kpc 
(Bloser et al. 2000). We assume $D = 8$~kpc for both GX 17+2 and 
GX 9+1, as their locations are believed to be near the galactic centre 
(Deutsch et al. 1999; Hertz et al. 1990) and distance of the galactic
centre is $7.9 \pm 0.3$ kpc, as concluded by McNamara et al. (2000).
For GX 349+2, we take $D = 9$~kpc (Deutsch et al. 1999).

We display the constrained values with the help of four tables. 
It is to be noted that here $\dot M$ is presented
in unit of $\dot M_{\rm e} = 1.4\times 10^{17} M/$\msun~g~s$^{-1}$. The 
Eddington rate is $\dot M_{\rm e}/\eta$, with  $\eta = E_{\rm BL}+E_{\rm D}$. 
Therefore, as the actual value of $\eta$ is much lesser than
1.0 (generally not greater than 0.3 and for rapidly rotating neutron star,
typically less than 0.2), the value of Eddington accretion rate is 
much higher than $\dot M_{\rm e}$. 
For all the sources, as the stiffness of the EOS models increases,
the absolute values of the allowed spin frequencies $(\nu_*)$ and rotational
frequencies in the ISCO $(\nu_{\rm in})$ decreases. This is because, for a
stiffer EOS model, neutron star radius is higher and it can support lesser 
amount of rotation. The energy conversion efficiency is also lesser 
for a stiffer EOS model (as the neutron star for this case is less compact)
and therefore higher accretion rate is needed to generate the observed 
luminosity (as seen from the tables). In the following, we describe the 
results for four sources in four subsections and give a general discussion
in the last subsection.

\subsection{XB 1820-30}

We display the allowed ranges of different parameters for the source 
XB 1820-30 in Table 2. It is seen that for $\cos i = 0.2$, the spin
frequency $(\nu_*)$ of the neutron star comes out to be very high. But in the 
case of $\cos i = 0.8$, it is not possible to constrain $(\nu_*)$ for 
(20\%,50\%) uncertainty set (for all EOS models) and for both the uncertainty 
sets (for EOS model D). The ranges of the colour factor are in general 
consistent with the results of Shimura \& Takahara $(f \sim 1.7 - 2.0)$ 
and Borozdin et al. (1999) $(f = 2.6)$. However, some discrepancy can be
noted with the latter one for softer EOS models \& $\cos i = 0.2$. 
The value of the rotational frequency in the ISCO $(\nu_{\rm in})$ comes out
to be $\sim 1$ kHz for all the cases. The values of the stellar equatorial
radius are in the range $8 - 21$ km. The peak of the disk effective 
temperature occurs in the radial range $18-30$ km, and always well-outside 
(by several kilometers) the neutron star's surface. This shows the validity
of eq. (4), as discussed in section 2. The overall range of the accretion rate 
comes out to be $0.5-31.4 \dot M_{\rm e}$.

\subsection{GX 17+2}

Table 3 shows the results for the source GX 17+2. Here the ranges of $(\nu_*)$
are similar to those for XB 1820-30. But for GX 17+2, the value of $i$ is
expected to be moderately high (Titarchuk et al. 2001) 
and therefore $\cos i$ is not possibly as high as 0.8. It is, therefore, 
quite likely that the neutron star in this source is rapidly rotating, 
although no decisive statement can be made. The allowed values for $f$
for GX 17+2 is systematically lower than those for XB 1820-30, and in the 
case of softer 
EOS models \& $\cos i = 0.2$ they do not tally with the result of Borozdin 
et al. (1999). The allowed values of $(\nu_{\rm in})$ coms out to be
$\sim 1$ kHz, but for EOS model (A) \& $\cos i = 0.8$, 2 kHz value is also
possible. The ranges of $R$ and $r^{\rm max}_{\rm eff}$ are similar to
those for XB 1820-30 and the allowed values of the accretion rate are in 
the range $2.0 - 131.0 \dot M_{\rm e}$.

\subsection{GX 9+1}

The results for the source GX 9+1 are given in Table 4. For this source,
$\nu_*$-value comes out to be very high for all the EOS models and for 
both the $\cos i$-values. Here, the allowed values for $f$ are inconsistent
with Borozdin et al. (1999) for softer EOS models \& $\cos i = 0.2$.
The allowed values of $\nu_{\rm in}$ are $\sim 1$ kHz and the allowed
ranges of $R$ and $r^{\rm max}_{\rm eff}$ are $10-21$ km and $18-30$ km 
respectively. The allowed values of accretion rate for this source 
come out to be in the range $3.8-116.7 \dot M_{\rm e}$.

\subsection{GX 349+2}

The allowed ranges of different parameters for the source GX 349+2 are 
given in Table 5.
As is the case for GX 9+1, here also the value of $\nu_*$ comes out 
to be very high for all the chosen EOS models and $\cos i$-values.
The allowed values for $f$ in general tally with the results of 
Shimura \& Takahara (1995), but like other three sources do not match
with the result of Borozdin et al. (1999) for softer EOS models \& 
$\cos i = 0.2$. Here $\nu_{\rm in}$ comes out to be $\sim 1$ kHz,
and $R$ \& $r^{\rm max}_{\rm eff}$ are in the ranges $10-21$ km and $18-30$ km
respectively (like GX 9+1). The accretion rate for this source 
is in the range $4.5-168.5 \dot M_{\rm e}$.

\subsection{General Discussion}

Here we have constrained the values of several properties of 
four LMXB sources. For all of them, the accretion rates come out to
be very high (always $\ge 0.5 ~\dot M_{\rm e})$. This is 
in accord with the fact that these are very luminous sources. 

The rotation rate of neutron star in each of the sources is very 
high (close to the mass shed limit) for $\cos i = 0.2$. This is 
because, the values of $L_{\rm BL}/L_{\rm D}$ are very low for these 
cases (see Thampan \& Datta 1998, Bhattacharyya et al. 2000). 
But, for $\cos i = 0.8$, rotation rate can not be constrained 
effectively for the sources XB 1820-30 and GX 17+2. Therefore, for 
these two sources, no general conclusion (about the values of 
$\Omega_{\rm *})$ can be drawn. However, 
the allowed ranges (combined for all the cases considered in a table) of
$\Omega_{\rm *}/\Omega_{\rm *,\mbox{ms}}$ are $0.93-1.00$ and $0.75-1.00$
for the other two sources GX 9+1 and GX 349+2 respectively
(here $\Omega_{\rm *,\mbox{ms}}$ is the $\Omega_{\rm *}$ at the mass
shed limit; see Bhattacharyya et al. 2000 for the mass shed limit values).
Therefore the neutron stars in these two sources
can be concluded to be rapidly rotating in general.

Our calculated allowed ranges for $f$ are in accord with the results 
obtained by Shimura \& Takahara (1995). However, 
if we take the value $f = 2.6$ (obtained by Borozdin et al. 1999), 
one would require a very stiff EOS model or a mass greater than 
$M = 1.4$~\msun for most of the cases with $\cos i = 0.2$.

High frequency quasi--periodic--oscillations (kHz QPO) have been 
observed for three (XB 1820-30, GX 17+2 and GX 349+2) of the chosen 
sources. The observed maximum kHz QPO frequencies are 1.100 kHz 
(XB 1820-30), 1.080 kHz (GX 17+2) and 1.020 kHz (GX 349+2) 
(van der Klis 2000). Now, as pointed out in Bhattacharyya et al. 
(2000), the maximum possible frequency (i.e., the shortest time scale)
of the system should be given by the rotational frequency in ISCO 
$(\nu_{\rm in}$; col. 5 of the tables). Therefore, the stiffest EOS 
model D is unfavoured for $\cos i = 0.2$ for the source XB 1820-30, 
as the maximum value of $\nu_{\rm in}$ $(= 0.941$~kHz, Table 2) 
is less than the observed maximum kHz QPO frequency. For the same 
reason, EOS model D is unfavoured for $\cos i 
= 0.2$ for the source GX 17+2. It can also be seen from Table 3 that 
if we use only the narrower limits on the luminosities and colour  
temperature, EOS model D (for $\cos i = 0.8)$ and EOS model C 
(for $\cos i = 0.2)$ are unfavoured for the same source. Same is true
for EOS model D for the source GX 349+2. As we also see from Table 5, 
EOS model C is unfavoured for $\cos i = 0.2$ for this source. 
Therefore, we may conclude that the stiffer EOS models are unfavoured 
by our results.

We have ignored the magnetic fields of the neutron stars in our
calculations. Therefore, the necessary condition for the validity of
our results is that the Alfv\'{e}n radius $(r_{\rm A})$ be less than the 
radius of the inner edge of the disc. This condition will always
be valid if $R > r_{\rm A}$ holds. Here $r_{\rm A}$ is given by
(Shapiro \& Teukolsky 1983),

\begin{eqnarray}
r_{\rm A} & = & 2.9 \times 10^8  {({\dot {M}\over \dot
{M}_{\rm e}})}^{-2/7} \mu_{30}^{4/7} ({M\over M_\odot})^{-3/7}
\end{eqnarray}

\noindent where
$M$ is the mass of the neutron star, $\mu_{30}$ is
the magnetic moment in units of $10^{30}$ G cm$^3$ and $r_{\rm A}$ is in cm.
With typical values of the parameters for the chosen sources $(R = 10$~km, 
$M = 1.4 M_\odot$ and $\dot{M} = 10 \dot {M}_{\rm e})$, the upper
limit of the neutron star surface magnetic field comes out to be
about $2 \times 10^{8}$~G. Therefore, our results are in general valid for 
the neutron star magnetic field upto of the order of $10^{8}$~G. This 
is a reasonable value for the magnetic field of neutron stars in LMXBs, as 
mentioned in section 1.

It is also to be noted that our results are valid for a thin blackbody
disc. However, as the spectra of the sources were well-fitted by a 
multicolour blackbody (plus a blackbody, presumably coming from the boundary 
layer; White et al. 1988), the assumption of thin blackbody disc may
be correct.

\section{Summary}

In this paper, we have constrained the values of two neutron star
parameters (spin frequency and equatorial radius) for the four
LMXBs: XB 1820-30, GX 17+2, GX 9+1 and GX 349+2. We have also 
calculated the allowed ranges of the colour factor (for accretion
disc), rotational frequency $(\nu_{\rm in})$ 
of a particle in the ISCO, the radius $(r^{\rm max}_{\rm eff})$
where the effective disc temperature is maximum and the accretion
rate for these sources. These have been done for a chosen mass $1.4$~\msun
(canonical mass) of the neutron stars and two values of 
inclination angles $(\cos i = 0.2, 0.8)$. The whole work has been repeated
for four EOS models (from very soft to very stiff).

We have drawn the following main conclusions from our study.
A comparison between the kHz QPO frequencies 
(observed from the sources) and our calculated
values of $\nu_{\rm in}$ has shown that the
stiffer EOS models are unfavoured. By the constraining procedure, 
we have got very high accretion rates for all the sources, which is
in accordance with their high luminosities. The neutron stars in the 
sources GX 9+1 and GX 349+2 have been found to be very rapidly
rotating, and those in the other two sources may also be rapidly rotating
(although we can not say decisively). This is in accordance with 
the belief that LMXBs are the progenitors of millisecond pulsars.
It also shows that while calculating the spectral models, taking 
the rapid rotation of neutron stars into account is very important.

It is difficult to constrain EOS models effectively with the 
poor quality {\it EXOSAT} data. However, the present generation
X--ray satellites {\it Chandra} and {\it XMM} have much better
resolving power. For example, the resolving power of {\it Chandra 
HETGS} is $60-1000$ in the energy range $0.5-10.0$ keV. 
The future generation X--ray satellite {\it Constellation-X} will 
have even better resolving power (upto 3000).
With the spectral data of these X--ray observatories, it may be 
possible to constrain EOS models and other parameters of LMXBs
effectively. Therefore we propose that it is essential to compute 
EOS dependent general relativistic spectral models to utilise 
these good quality data in a fruitful way. 

\acknowledgements

We acknowledge Arun Thampan for providing us with the 
neutron star structure
calculation code and for discussions. 
We thank Dipankar Bhattacharya 
for reading the manuscript and giving valuable suggestions. 
We also thank Ranjeev Misra and late Bhaskar Datta for discussions and 
Pijush Bhattacharjee for encouragement.

\appendixname

Here we give the Einstein's field equations and the equation of hydrostatic
equilibrium, that were solved for the computation of the structure of the
rapidly rotating neutron star.
For an axisymmetric and equatorial plane symmetric configuration, the
computational domain in spherical polar coordinates covers
$0 \le r \le \infty$ and $0 \le \theta \le \pi/2$. For numerical convenience,
we make a change of variables ($r \rightarrow s$ and $\theta \rightarrow
\mu$) given by
{\bf $\tilde{r} = \tilde{r_{\rm e}} \frac{s}{1-s}$ and
$\theta = \cos^{-1} \mu$,}
where $\bar{r_{\rm e}}$ is the quasi--isotropic radial 
coordinate of the
equator. It is easy to see that $s$ and $\mu$ vary in the range
$0 \le s \le 1$ \& $0 \le \mu \le 1$ and at the equator $s = 0.5$.

The four Einstein's equations (Cook et al. 1994) to solve are given below:

\begin{eqnarray}
\rho(s, \mu) & = & -e^{-\gamma/2} \sum_{n=0}^{\infty} P_{2 n}(\mu)
\mbox{\Huge [}\mbox{\Huge (}\frac{1-s}{s}\mbox{\Huge )}^{2 n+1} \int_0^s
\frac{ds' s'^{2 n}}{(1-s')^{2 n+2}} \nonumber \\
 &   & \times \int_0^1 d\mu' P_{2 n}(\mu')
\tilde S_{\rho}(s', \mu') 
+ \mbox{\Huge (}\frac{s}{1-s}\mbox{\Huge )}^{2 n} \nonumber \\
 &   & \times \int_s^1
\frac{ds' (1-s')^{2 n-1}}{s'^{2 n+1}} 
\int_0^1 d\mu' P_{2 n}(\mu')
\tilde S_{\rho}(s', \mu')\mbox{\Huge ]}
\end{eqnarray}

\begin{eqnarray}
\gamma(s, \mu) & = & -\frac{2 e^{-\gamma/2}}{\pi} \sum_{n=1}^{\infty}
\frac{\sin[(2 n-1) \theta]}{(2 n-1) \sin \theta}
\mbox{\Huge [}\mbox{\Huge (}\frac{1-s}{s}\mbox{\Huge )}^{2 n} \nonumber \\
 &   & \times \int_0^s \frac{ds' s'^{2 n-1}}{(1-s')^{2 n+1}}
\int_0^1 d\mu' \sin[(2 n-1) \theta'] \tilde S_{\gamma}(s', \mu') \nonumber \\
 &   & + \mbox{\Huge (}\frac{s}{1-s}\mbox{\Huge )}^{2 n-2} 
\int_s^1 \frac{ds' (1-s')^{2 n-3}}{s'^{2 n-1}} \nonumber \\
 &   & \times \int_0^1 d\mu' \sin[(2 n-1) \theta']
\tilde S_{\gamma}(s', \mu')\mbox{\Huge ]}
\end{eqnarray}

\begin{eqnarray}
\hat \omega(s, \mu) & = & -e^{(2 \rho-\gamma)/2} \sum_{n=1}^{\infty}
\frac{P^1_{2 n-1}(\mu)}{2 n (2 n-1) \sin \theta} \mbox{\Huge [}\mbox{\Huge
(}\frac{1-s}{s}\mbox{\Huge )}^{2 n+1} \nonumber \\
 &   & \times \int_0^s \frac{ds' s'^{2 n}}{(1-s')^{2 n+2}}
\int_0^1 d\mu' \sin \theta' P^1_{2 n-1}(\mu')
\tilde S_{\hat \omega}(s', \mu') \nonumber \\
 &   & + \mbox{\Huge
(}\frac{s}{1-s}\mbox{\Huge )}^{2 n-2} 
\int_s^1 \frac{ds' (1-s')^{2 n-3}}{s'^{2 n-1}} \nonumber \\
 &   & \times \int_0^1 d\mu' \sin \theta' P^1_{2 n-1}(\mu')
\tilde S_{\hat \omega}(s', \mu')\mbox{\Huge ]}
\end{eqnarray}

\begin{eqnarray}
\alpha_{,\mu} & = & -\frac{1}{2} (\rho_{,\mu} + \gamma_{,\mu}) -
\{(1-\mu^2) [1+s (1-s) \gamma_{,s}]^2 \nonumber \\
 &   & + [-\mu+(1-\mu^2)\gamma_{,\mu}]^2\}^{-1} 
\mbox{\Huge [}\frac{1}{2} \{s (1-s) [s (1-s) \gamma_{,s}]_{,s} \nonumber \\
 &   & + s^2 (1-s)^2 \gamma_{,s}^2 - [(1-\mu^2) 
\gamma_{,\mu}]_{,\mu} \nonumber \\
 &   & - \gamma_{,\mu} [-\mu+(1-\mu^2) \gamma_{,\mu}]\}
[-\mu+(1-\mu^2) \gamma_{,\mu}] \nonumber \\ 
 &   & + \frac{1}{4} [s^2 (1-s)^2 (\rho_{,s}+
\gamma_{,s})^2 
- (1-\mu^2) (\rho_{,\mu}+\gamma_{,\mu})^2] \nonumber \\
 &   & \times [-\mu+(1-\mu^2) \gamma_{,\mu}] - s (1-s) (1-\mu^2) \nonumber \\
 &   & \times \mbox{\Huge (}\frac{1}{2} (\rho_{,s}+\gamma_{,s})
(\rho_{,\mu}+\gamma_{,\mu}) + \gamma_{,s\mu} + \gamma_{,s}
\gamma_{,\mu}\mbox{\Huge )} [1+s (1-s) \gamma_{,s}] \nonumber \\
 &   & + s (1-s) \mu \gamma_{,s} [1+s (1-s) \gamma_{,s}] + \frac{1}{4}
(1-\mu^2) e^{-2 \rho} \nonumber \\
 &   & \times \mbox{\Huge \{}2 \frac{s^3}{1-s} (1-\mu^2) \hat{\omega}_{,s}
\hat{\omega}_{,\mu} [1+s (1-s) \gamma_{,s}] \nonumber \\
 &   & - \mbox{\Huge (}s^4 \hat{\omega}_{,s}^2 - \frac{s^2}{(1-s)^2}
(1-\mu^2) \hat{\omega}_{,s}^2\mbox{\Huge )} [-\mu \nonumber \\
 &   & +(1-\mu^2)
\gamma_{,\mu}]\mbox{\Huge \}]}
\end{eqnarray}

\noindent 
where $P_n(\mu)$ are the Legendre polynomials, $P_n^m(\mu)$
are the associated Legendre polynomials and $\sin(n\theta)$ is a function of
$\mu$ through $\theta = \cos^{-1} \mu$.
The effective sources $\tilde S$'s are defined as (Cook et al. 1994)

\begin{eqnarray}
\tilde S_{\rho}(s, \mu) & = & e^{\gamma/2} \mbox{\Huge [}8 \pi e^{2 \alpha}
\tilde r_{\rm e}^2 (\tilde \epsilon + \tilde P)
\mbox{\Huge (}\frac{s}{1-s}\mbox{\Huge )}^2
\frac{1+\tilde v^2}{1-\tilde v^2} \nonumber \\
 &   & + \mbox{\Huge (}\frac{s}{1-s}\mbox{\Huge )}^2 (1-\mu^2) e^{-2 \rho}
\{[s (1-s) \hat \omega_{,s}]^2 + (1-\mu^2) \hat \omega_{,\mu}^2\} \nonumber \\
 &   & + s (1-s) \gamma_{,s} - \mu \gamma_{,\mu} + \frac{\rho}{2}
\mbox{\Huge \{}16 \pi e^{2 \alpha} \tilde r_{\rm e}^2 \tilde P
\mbox{\Huge (}\frac{s}{1-s}\mbox{\Huge )}^2 \nonumber \\
 &   & - s (1-s) \gamma_{,s} \mbox{\Huge (}\frac{s (1-s)}{2} \gamma_{,s}
+ 1\mbox{\Huge )} \nonumber \\ 
 &   & - \gamma_{,\mu} \mbox{\Huge (}\frac{1-\mu^2}{2}
\gamma_{,\mu} - \mu\mbox{\Huge )\}]}
\end{eqnarray}

\begin{eqnarray}
\tilde S_{\gamma}(s, \mu) & = & e^{\gamma/2} \mbox{\Huge [}16 \pi e^{2 \alpha}
 \tilde r_{\rm e}^2 \tilde P\mbox{\Huge (}\frac{s}{1-s}\mbox{\Huge )}^2 
+ \frac{\gamma}{2} \mbox{\Huge \{}16 \pi e^{2 \alpha} \tilde r_{\rm e}^2
\tilde P\mbox{\Huge (}\frac{s}{1-s}\mbox{\Huge )}^2 \nonumber \\
 &   &  - \frac{s^2 (1-s)^2}{2} \gamma_{,s}^2 - \frac{1-\mu^2}{2}
\gamma_{,\mu}^2\mbox{\Huge \}]}
\end{eqnarray}

\begin{eqnarray}
\tilde S_{\hat{\omega}}(s, \mu) & = & e^{(\gamma-2 \rho)/2}
\mbox{\Huge [}-16 \pi e^{2 \alpha} \frac{(\hat \Omega_* - \hat \omega)}{1 -
\tilde v^2} \tilde r_{\rm e}^2 (\tilde \epsilon + \tilde P) \mbox{\Huge
(}\frac{s}{1-s}\mbox{\Huge )}^2 \nonumber \\
 &   & + \hat \omega \mbox{\Huge \{}-8 \pi e^{2 \alpha} \tilde r_{\rm e}^2
\frac{(1+\tilde v^2) \tilde \epsilon + 2 \tilde v^2 \tilde P}{1-\tilde v^2}
\mbox{\Huge (}\frac{s}{1-s}\mbox{\Huge )}^2 \nonumber \\
 &   & - s (1-s) \mbox{\Huge (}2 \rho_{,s} + \frac{1}{2}
\gamma_{,s}\mbox{\Huge )} + \mu \mbox{\Huge (}2 \rho_{,\mu} + \frac{1}{2}
\gamma_{,\mu}\mbox{\Huge )} \nonumber \\
 &   & + \frac{s^2 (1-s)^2}{4} (4 \rho_{,s}^2 - \gamma_{,s}^2) +
\frac{1-\mu^2}{4} (4 \rho_{,\mu}^2 - \gamma_{,\mu}^2) \nonumber \\
 &   & - (1-\mu^2) e^{-2 \rho} \mbox{\Huge (}s^4 \hat{\omega}_{,s}^2 +
\frac{s^2 (1-\mu^2)}{(1-s)^2} \hat{\omega}_{,\mu}^2\mbox{\Huge )\}]}
\end{eqnarray}

{\bf\noindent where
$\hat \omega \equiv  \tilde r_{\rm e} \tilde \omega$ and
$\hat \Omega_* \equiv  \tilde r_{\rm e} \tilde \Omega_*$.}
Here the tilde over a variable represents the corresponding
dimensionless quantity and the variables $\epsilon$ and $P$ 
(not to be confused with Legendre polynomials, that always
have a function of `n' as the subscript) represent mass--energy density
and pressure respectively.

The equation of hydrostatic equilibrium for a barytropic fluid is

\begin{eqnarray}
h(\tilde P) - h_{\rm p} & \equiv & \int_{\tilde P_{\rm p}}^{\tilde P}
\frac{d\tilde P}{\tilde \epsilon + \tilde P} \nonumber \\
 & = & \ln u^t - \ln u_{\rm p}^t
- \int_{\tilde \Omega_{\rm *, c}}^{\tilde \Omega_*} F(\tilde \Omega_*) d\tilde \Omega_*
\end{eqnarray}

\noindent where $h(\tilde P)$ is the dimensionless specific enthalpy as a
function of pressure and $\tilde P_{\rm p}, u_{\rm p}^t$ and $h_{\rm p}$ are
the dimensionless values of pressure, t-component of the four--velocity and the
specific enthalpy at the pole. The quantity
$\tilde \Omega_{\rm *, c}$ is the (dimensionless)
central value of the angular speed, which on the rotation axis is constant
and equal to its value at the pole. The quantity 
$F(\tilde \Omega_*) = u^t u_{\phi}$
is obtained from an integrability condition on the equation of
hydrostatic equilibrium. Choosing the form of this function fixes the
rotation law for the matter. Following Komatsu et al. (1989), we set it to
{\bf $F(\tilde \Omega_*) = A^2 (\tilde \Omega_{\rm *, c} - \tilde \Omega_*)$.
Here} $A$ is a rotation constant such that rigid rotation is
achieved in the limit $A \rightarrow \infty$. An appropriately chosen value
of $h_{\rm p}$ defines the surface of the star.

\begin{table*}
\caption{Best-fit values of the parameters (see text) for four LMXBs.}
\begin{tabular}{ccccc}
\hline
\hline
\multicolumn{1}{c}{Source name} & \multicolumn{1}{c}{$T_{\rm col}^{\rm max}$}  &
\multicolumn{2}{c}{$L_{\rm D}$} & \multicolumn{1}{c}{$L_{\rm BL}$} \\
\multicolumn{1}{c}{}  
& \multicolumn{1}{c}{$(10^7$ K)}  & \multicolumn{2}{c}{$(10^{38}$ ergs s$^{-1})$} 
& \multicolumn{1}{c}{$(10^{38}$ ergs s$^{-1})$} \\
 & & $\cos i = 0.2$ & $\cos i = 0.8$ & \\
\hline
XB 1820-30 & 1.59 & 1.49 & 0.37 & 0.26 \\
GX 17+2 & 1.76 & 6.49 & 1.62 & 0.71 \\
GX 9+1 & 2.25 & 6.01 & 1.50 & 0.25 \\
GX 349+2 & 2.07 & 8.54 & 2.14 & 0.48 \\
\hline
\end{tabular}
\end{table*}


\begin{table*}
\caption{Observational constraints for various EOS models : (A),
(B), (C), (D) for the source XB 1820-30. L and U stand 
for lower and upper limits. The parameters
are $f$ 
(colour factor), $\nu_{\ast}$ (rotational frequency of the neutron star),
$\nu_{\rm in}$ (rotational frequency in the ISCO),
$R$ (equatorial radius of the neutron star), $r_{\rm eff}^{\rm max}$
(radius where the effective temperature of the disc is maximum) and
$\dot {M}$ (the accretion rate).  The limits are for 25\% uncertainty in
luminosities and 10\% uncertainty in the colour temperature.
Values in brackets are for 50\% uncertainty in luminosities and 20\% uncertainty
in the colour temperature. $i$ is the inclination angle of the source 
with respect to the observer. The mass of the neutron star is assumed to be
1.4 $M_\odot$. The accretion rate is given in unit of $\dot {M}_{\rm e} = 
1.4\times 10^{17} M/$\msun~g~s$^{-1}$, where $M$ is the neutron star mass.}

\begin{tabular}{lllllllll}
\hline
\hline
EOS  & $\cos i$ &  & $f$  & $\nu_{\ast}$& $\nu_{\rm in}$ & $R$ &
$r_{\rm eff}^{\rm max}$ & $\dot{M}$   \\
   & &  & &  kHz & kHz & km & km & $\dot{M}_{\rm e}$  \\
\hline
(A)  &   0.2 & L  & 1.31[1.10] & 1.751[1.726] & 1.756[1.756] & 11.2[10.2] & 
18.6[18.1] & 7.7[3.8]  \\
  &  &       U  & 1.91[2.41] & 1.755[1.755] & 1.819[2.078] & 11.4[11.4] & 
18.7[18.7] & 16.1[20.3]  \\
\hline
(B)  &   0.2 & L  & 1.45[1.30] & 1.103[1.059] & 1.137[1.126] & 15.0[13.7] & 
23.0[21.6] & 9.7[4.5] \\
  &  &       U  & 2.07[2.61] & 1.112[1.113] & 1.197[1.372] & 15.5[15.6] & 
23.5[23.6] & 19.4[24.4]  \\
\hline
(C)  &   0.2 & L  & 1.49[1.30] & 0.961[0.913] & 0.979[0.973] & 16.5[15.0] & 
24.8[23.1] & 10.4[4.9]  \\
  &  &       U  & 2.12[2.71] & 0.968[0.968] & 1.042[1.206] & 17.2[17.2] & 
25.5[25.6] & 21.2[26.7]  \\
\hline
(D)  &   0.2 & L  & 1.59[1.40] & 0.735[0.687] & 0.748[0.743] & 19.9[17.7] & 
29.1[26.5] & 12.2[5.6]  \\
  &  &       U  & 2.25[2.81] & 0.740[0.740] & 0.795[0.941] & 20.6[20.7] & 
30.0[30.1] & 25.0[31.4]  \\
\hline
(A)  &   0.8 & L  & 1.79[1.50] & 1.463[0.000] & 1.822[1.571] & 9.9[7.5] & 
18.1[18.1] & 1.2[0.5]  \\
  &  &       U  & 3.06[4.70] & 1.751[1.754] & 2.165[2.165] & 11.2[11.4] & 
20.4[22.3] & 4.5[6.4]  \\
\hline
(B)  &   0.8 & L  & 1.94[1.71] & 0.498[0.000] & 1.207[1.152] & 11.3[11.0] & 
20.2[20.2] & 1.4[0.9]  \\
  &  &        U  & 3.22[4.20] & 1.102[1.110] & 1.782[1.782] & 14.9[15.4] & 
22.9[23.4] & 5.6[7.9]  \\
\hline
(C)  &   0.8 & L  & 1.99[1.70] & 0.175[0.000] & 1.046[0.991] & 12.3[12.3] & 
21.5[21.0] & 1.5[1.0]  \\
  &  &        U  & 3.36[4.02] & 0.960[0.966] & 1.573[1.573] & 16.5[17.0] & 
24.7[25.4] & 6.1[8.5]  \\
\hline
(D)  &   0.8 & L  & 2.11[1.80]  & 0.000[0.000] & 0.806[0.758] & 14.7[14.7] & 
23.1[23.1] & 1.8[1.2]  \\
  &  &        U  & 3.30[3.90]  & 0.733[0.739] & 1.212[1.212] & 19.7[20.5] & 
28.9[29.8] & 7.2[9.9]  \\
\hline
\end{tabular}
\end{table*}


\begin{table*}
\caption{Observational constraints for various EOS models : (A),
(B), (C), (D) for the source GX 17+2. 
Other specifications are same as in Table 2.}

\begin{tabular}{lllllllll}
\hline
\hline
EOS  & $\cos i$ &  & $f$  & $\nu_{\ast}$& $\nu_{\rm in}$ & $R$ &
$r_{\rm eff}^{\rm max}$ & $\dot{M}$   \\
   & &  & &  kHz & kHz & km & km & $\dot{M}_{\rm e}$  \\
\hline
(A)  &   0.2 & L  & 1.01[1.00] & 1.754[1.748] & 1.756[1.756] & 11.4[11.0] & 
18.7[18.5] & 36.1[19.8]  \\
  &  &       U  & 1.43[1.82] & 1.755[1.755] & 1.773[1.869] & 11.4[11.4] & 
18.7[18.7] & 68.7[82.6]  \\
\hline
(B)  &   0.2 & L  & 1.12[1.00] & 1.108[1.097] & 1.128[1.122] & 15.3[14.7] & 
23.3[22.6] & 44.4[24.4] \\
  &  &       U  & 1.57[1.96] & 1.113[1.113] & 1.163[1.236] & 15.6[15.7] & 
23.6[23.6] & 82.6[101.7]  \\
\hline
(C)  &   0.2 & L  & 1.15[1.00] & 0.966[0.954] & 0.974[0.971] & 16.9[16.0] & 
25.2[24.2] & 47.5[26.1]  \\
  &  &       U  & 1.62[2.01] & 0.968[0.968] & 1.000[1.091] & 17.2[17.2] & 
25.5[25.6] & 88.5[111.5]  \\
\hline
(D)  &   0.2 & L  & 1.22[1.03] & 0.738[0.728] & 0.744[0.742] & 20.3[19.2] & 
29.7[28.3] & 55.9[30.7]  \\
  &  &       U  & 1.72[2.13] & 0.740[0.740] & 0.766[0.849] & 20.7[20.7] & 
30.1[30.1] & 106.4[131.0]  \\
\hline
(A)  &   0.8 & L  & 1.39[1.20] & 1.702[0.000] & 1.782[1.571] & 9.9[7.5] & 
18.1[18.1] & 6.6[2.0]  \\
  &  &       U  & 2.20[3.72] & 1.754[1.755] & 2.166[2.166] & 11.3[11.4] & 
18.6[22.3] & 18.5[25.5]  \\
\hline
(B)  &   0.8 & L  & 1.53[1.31] & 1.009[0.000] & 1.172[1.141] & 13.1[11.0] & 
21.1[20.2] & 7.7[3.2]  \\
  &  &        U  & 2.31[3.35] & 1.107[1.111] & 1.463[1.782] & 15.2[15.5] & 
23.2[23.5] & 22.8[30.7]  \\
\hline
(C)  &   0.8 & L  & 1.61[1.30] & 0.858[0.000] & 1.010[0.983] & 14.3[12.3] & 
22.4[21.0] & 8.3[3.6]  \\
  &  &        U  & 2.33[3.20] & 0.965[0.967] & 1.289[1.568] & 16.8[17.1] & 
25.1[25.4] & 25.0[32.9]  \\
\hline
(D)  &   0.8 & L  & 1.66[1.41]  & 0.631[0.000] & 0.775[0.750] & 16.8[14.7] & 
25.4[23.1] & 9.5[4.4]  \\
  &  &        U  & 2.47[3.10]  & 0.737[0.740] & 1.011[1.212] & 20.2[20.6] & 
29.5[30.0] & 29.3[38.6]  \\
\hline
\end{tabular}
\end{table*}


\begin{table*}
\caption{Observational constraints for various EOS models : (A),
(B), (C), (D) for the source GX 9+1. 
Other specifications are same as in Table 2.}

\begin{tabular}{lllllllll}
\hline
\hline
EOS  & $\cos i$ &  & $f$  & $\nu_{\ast}$& $\nu_{\rm in}$ & $R$ &
$r_{\rm eff}^{\rm max}$ & $\dot{M}$   \\
   & &  & &  kHz & kHz & km & km & $\dot{M}_{\rm e}$  \\
\hline
(A)  &   0.2 & L  & 1.33[1.13] & 1.755[1.755] & 1.756[1.756] & 11.4[11.4] &
18.7[18.7] & 36.9[22.8]  \\
  &  &       U  & 1.85[2.25] & 1.755[1.755] & 1.756[1.761] & 11.4[11.4] &
18.7[18.7] & 59.9[72.0]  \\
\hline
(B)  &   0.2 & L  & 1.47[1.24] & 1.112[1.110] & 1.120[1.117] & 15.6[15.4] &
23.6[23.4] & 43.4[27.4] \\
  &  &       U  & 2.04[2.49] & 1.114[1.114] & 1.130[1.147] & 15.7[15.7] &
23.6[23.7] & 73.6[90.6]  \\
\hline
(C)  &   0.2 & L  & 1.51[1.28] & 0.968[0.967] & 0.970[0.970] & 17.2[17.1] &
25.5[25.4] & 47.5[29.3]  \\
  &  &       U  & 2.09[2.56] & 0.968[0.968] & 0.975[0.986] & 17.3[17.3] &
25.6[25.6] & 80.7[99.3]  \\
\hline
(D)  &   0.2 & L  & 1.61[1.36] & 0.740[0.739] & 0.742[0.741] & 20.7[20.5] &
30.1[29.9] & 55.9[34.4]  \\
  &  &       U  & 2.24[2.74] & 0.740[0.740] & 0.745[0.754] & 20.7[20.7] &
30.1[30.1] & 94.9[116.7]  \\
\hline
(A)  &   0.8 & L  & 1.84[1.61] & 1.752[1.728] & 1.756[1.756] & 11.2[10.3] &
18.6[18.1] & 7.9[3.8]  \\
  &  &       U  & 2.69[3.52] & 1.755[1.755] & 1.815[2.064] & 11.4[11.4] &
18.7[18.7] & 16.5[20.3]  \\
\hline
(B)  &   0.8 & L  & 2.05[1.80] & 1.103[1.064] & 1.136[1.126] & 15.0[13.8] &
22.9[21.7] & 9.7[4.6]  \\
  &  &        U  & 2.92[3.72] & 1.112[1.113] & 1.200[1.361] & 15.5[15.6] &
23.5[23.6] & 19.4[24.4]  \\
\hline
(C)  &   0.8 & L  & 2.10[1.80] & 0.961[0.919] & 0.978[0.972] & 16.5[15.0] &
24.8[23.2] & 10.4[5.0]  \\
  &  &        U  & 3.00[3.81] & 0.968[0.968] & 1.041[1.195] & 17.2[17.2] &
25.5[25.6] & 21.2[26.7]  \\
\hline
(D)  &   0.8 & L  & 2.24[1.92]  & 0.734[0.692] & 0.748[0.743] & 19.8[17.8] &
29.0[26.7] & 12.2[5.7]  \\
  &  &        U  & 3.19[4.00]  & 0.740[0.740] & 0.802[0.932] & 20.6[20.7] &
30.0[30.1] & 25.0[31.4]  \\
\hline
\end{tabular}
\end{table*}


\begin{table*}
\caption{Observational constraints for various EOS models : (A),
(B), (C), (D) for the source GX 349+2. 
Other specifications are same as in Table 2.}

\begin{tabular}{lllllllll}
\hline
\hline
EOS  & $\cos i$ &  & $f$  & $\nu_{\ast}$& $\nu_{\rm in}$ & $R$ &
$r_{\rm eff}^{\rm max}$ & $\dot{M}$   \\
   & &  & &  kHz & kHz & km & km & $\dot{M}_{\rm e}$  \\
\hline
(A)  &   0.2 & L  & 1.12[1.00] & 1.755[1.754] & 1.756[1.756] & 11.4[11.4] &
18.7[18.7] & 50.9[30.7]  \\
  &  &       U  & 1.55[1.92] & 1.755[1.755] & 1.756[1.771] & 11.4[11.4] &
18.7[18.7] & 86.4[103.9]  \\
\hline
(B)  &   0.2 & L  & 1.24[1.04] & 1.112[1.109] & 1.122[1.119] & 15.5[15.3] &
23.5[23.3] & 61.2[37.7] \\
  &  &       U  & 1.72[2.11] & 1.113[1.114] & 1.137[1.159] & 15.7[15.7] &
23.6[23.7] & 106.3[130.8]  \\
\hline
(C)  &   0.2 & L  & 1.27[1.08] & 0.968[0.966] & 0.971[0.970] & 17.2[17.0] &
25.5[25.3] & 67.1[40.4]  \\
  &  &       U  & 1.76[2.17] & 0.968[0.968] & 0.979[1.000] & 17.2[17.3] &
25.6[25.6] & 116.6[140.1]  \\
\hline
(D)  &   0.2 & L  & 1.35[1.14] & 0.740[0.738] & 0.742[0.741] & 20.6[20.4] &
30.0[29.7] & 78.8[47.5]  \\
  &  &       U  & 1.88[2.31] & 0.740[0.740] & 0.748[0.765] & 20.7[20.7] &
30.1[30.1] & 136.9[168.5]  \\
\hline
(A)  &   0.8 & L  & 1.55[1.34] & 1.748[1.678] & 1.760[1.756] & 11.0[9.7] &
18.4[18.1] & 10.6[4.5]  \\
  &  &       U  & 2.29[3.10] & 1.755[1.755] & 1.879[2.147] & 11.4[11.4] &
18.7[18.7] & 23.3[30.0]  \\
\hline
(B)  &   0.8 & L  & 1.71[1.50] & 1.096[0.955] & 1.148[1.129] & 14.6[12.7] &
22.6[20.7] & 13.1[5.2]  \\
  &  &        U  & 2.47[3.22] & 1.110[1.112] & 1.244[1.532] & 15.4[15.6] &
23.4[23.6] & 28.0[36.0]  \\
\hline
(C)  &   0.8 & L  & 1.76[1.53] & 0.953[0.798] & 0.984[0.974] & 16.0[13.8] &
24.2[21.9] & 14.0[5.6]  \\
  &  &        U  & 2.54[3.30] & 0.967[0.968] & 1.093[1.350] & 17.1[17.2] &
25.4[25.5] & 30.7[38.6]  \\
\hline
(D)  &   0.8 & L  & 1.87[1.61]  & 0.727[0.557] & 0.753[0.744] & 19.1[16.2] &
28.2[24.6] & 16.5[6.3]  \\
  &  &        U  & 2.69[3.40]  & 0.739[0.740] & 0.845[1.069] & 20.6[20.7] &
29.9[30.1] & 36.0[46.4]  \\
\hline
\end{tabular}
\end{table*}

\begin{thebibliography}{}

\bibitem[]{} Backer, D.C., Kulkarni, S.R., Heiles, C., Davis, M.M., Goss, W.M. 
1982, Nature, 300, 615

\bibitem[]{} Baldo, M., Bombaci, I., \& Burgio, G.F. 1997, A\&A, 328, 274

\bibitem[]{} Bhattacharya, D., \& Datta, B. 1996, MNRAS, 282, 1059

\bibitem[]{} Bhattacharya, D., \& van den Heuvel, E.P.J. 1991, Phys. Repts., 
203, 1

\bibitem[]{} Bhattacharyya, S., Thampan, A.V., Misra, R. \& Datta, B. 2000, 
ApJ, 542, 473

\bibitem[]{} Bloser, P.F., Grindlay, J.E., Kaaret, P., Zhang, W., Smale, A.P., 
\& Barret, D. 2000, ApJ, 542, 1000

\bibitem[]{} Borozdin, K., Revnivtsev, M., Trudolyubov, S., Shrader, C., \&
             Titarchuk, L. 1999, ApJ, 517, 367

\bibitem[]{} Cook, G.B., Shapiro, S.L., \& Teukolsky, S.A. 1994, ApJ, 424, 823

\bibitem[]{} Deutsch, E.W., Margon, B., Anderson, S.F., Wachter, S., \& 
Goss, W.M. 1999, ApJ, 524, 406

\bibitem[]{} Hanawa, T. 1989, ApJ, 341, 948

\bibitem[]{} Hertz, P., Norris, J.P., Wood, K.S., Vaughan, B.A. \& 
Michelson, P.F. 1990, ApJ, 354, 267

\bibitem[]{} Komatsu, H., Eriguchi, Y., \& Hachisu, I. 1989, MNRAS, 237, 355

\bibitem[]{} McNamara, D.H., Madsen, J.B., Barnes, J., \& Ericksen, B.F. 
2000, PASP, 112, 202 

\bibitem[]{} Mitsuda, K., Inoue, H., Koyama, K., Makishima, K.
et al. 1984, PASJ, 36, 741

\bibitem[]{} Page, D.N., \& Thorne, K.S. 1974, ApJ, 191, 499

\bibitem[]{} Pandharipande, V.R. 1971, Nucl. Phys. A, 178, 123

\bibitem[]{} Sahu, P.K., Basu, R., \& Datta, B. 1993, ApJ, 416, 267

\bibitem[]{} Shapiro, S.L., \& Teukolsky, S.A. 1983, in Black Holes, 
White Dwarfs, and Neutron Stars (New York : John Wiley \& Sons)

\bibitem[]{} Shimura, T., \& Takahara, F. 1995, ApJ, 445, 780

\bibitem[]{} Titarchuk, L.G., Bradshaw, C.F., Geldzahler, B.J., \& 
Fomalont, E.B. 2001, astro-ph/0105559

\bibitem[]{} Thampan, A.V., \& Datta, B. 1998, MNRAS, 297, 570

\bibitem[]{} van der Klis, M. 2000, ARAA, 38, 717

\bibitem[]{} van Paradijs, J., 1995, in X--Ray Binaries, ed. Lewin, W.H.G., 
van Paradijs, J., \& van den Heuvel, E.P.J., 536

\bibitem[]{} Walecka, J.D. 1974, Ann. Phys., 83, 491

\bibitem[]{} White, N.E., Stella, L., \& Parmar, A.N. 1988, ApJ, 324, 363

\bibitem[]{} Wijnands, R., \& van der Klis, M. 1998, Nature, 394, 344

\end{thebibliography}
\end{document}